\documentclass[11pt,a4paper]{article}
\usepackage{amsmath,amssymb,latexsym}
\usepackage{mathrsfs,graphicx,subfigure}

\def\AFOUR{%
\setlength{\textheight}{8.5in}%
\setlength{\textwidth}{5.75in}%
\setlength{\topmargin}{-0.375in}%
\hoffset=-.5in%
\renewcommand{\baselinestretch}{1.17}%
\setlength{\parskip}{6pt plus 2pt}%
}

\AFOUR                                           %%%   ***

\expandafter\ifx\csname amssym.def\endcsname\relax \else\endinput\fi
\expandafter\edef\csname amssym.def\endcsname{%
       \catcode`\noexpand\@=\the\catcode`\@\space}
\catcode`\@=11
\def\undefine#1{\let#1\undefined}
\def\newsymbol#1#2#3#4#5{\let\next@\relax
 \ifnum#2=\@ne\let\next@\msafam@\else
 \ifnum#2=\tw@\let\next@\msbfam@\fi\fi
 \mathchardef#1="#3\next@#4#5}
\def\mathhexbox@#1#2#3{\relax
 \ifmmode\mathpalette{}{\m@th\mathchar"#1#2#3}%
 \else\leavevmode\hbox{$\m@th\mathchar"#1#2#3$}\fi}
\def\hexnumber@#1{\ifcase#1 0\or 1\or 2\or 3\or 4\or 5\or 6\or 7\or 8\or
 9\or A\or B\or C\or D\or E\or F\fi}

\font\tenmsa=msam10
\font\sevenmsa=msam7
\font\fivemsa=msam5
\newfam\msafam
\textfont\msafam=\tenmsa
\scriptfont\msafam=\sevenmsa
\scriptscriptfont\msafam=\fivemsa
\edef\msafam@{\hexnumber@\msafam}
\mathchardef\dabar@"0\msafam@39
\def\dashrightarrow{\mathrel{\dabar@\dabar@\mathchar"0\msafam@4B}}
\def\dashleftarrow{\mathrel{\mathchar"0\msafam@4C\dabar@\dabar@}}

\def\ulcorner{\delimiter"4\msafam@70\msafam@70 }
\def\urcorner{\delimiter"5\msafam@71\msafam@71 }
\def\llcorner{\delimiter"4\msafam@78\msafam@78 }
\def\lrcorner{\delimiter"5\msafam@79\msafam@79 }
\def\yen{{\mathhexbox@\msafam@55}}
\def\checkmark{{\mathhexbox@\msafam@58}}
\def\circledR{{\mathhexbox@\msafam@72}}
\def\maltese{{\mathhexbox@\msafam@7A}}
\def\circledS{{\mathhexbox@\msafam@73}}

\font\tenmsb=msbm10
\font\sevenmsb=msbm7
\font\fivemsb=msbm5
\newfam\msbfam
\textfont\msbfam=\tenmsb
\scriptfont\msbfam=\sevenmsb
\scriptscriptfont\msbfam=\fivemsb
\edef\msbfam@{\hexnumber@\msbfam}
\def\Bbb#1{{\fam\msbfam\relax#1}}
\def\widehat#1{\setbox\z@\hbox{$\m@th#1$}%
 \ifdim\wd\z@>\tw@ em\mathaccent"0\msbfam@5B{#1}%
 \else\mathaccent"0362{#1}\fi}
\def\widetilde#1{\setbox\z@\hbox{$\m@th#1$}%
 \ifdim\wd\z@>\tw@ em\mathaccent"0\msbfam@5D{#1}%
 \else\mathaccent"0365{#1}\fi}
\font\teneufm=eufm10
\font\seveneufm=eufm7
\font\fiveeufm=eufm5
\newfam\eufmfam
\textfont\eufmfam=\teneufm
\scriptfont\eufmfam=\seveneufm
\scriptscriptfont\eufmfam=\fiveeufm
\def\frak#1{{\fam\eufmfam\relax#1}}

\csname amssym.def\endcsname

\makeatletter
\def\section{\@startsection {section}{1}{\z@}{-3.5ex plus -1ex minus
 -.2ex}{2.3ex plus .2ex}{\large\sc}}
\def\subsection{\@startsection{subsection}{2}{\z@}{-3.25ex plus -1ex minus
 -.2ex}{1.5ex plus .2ex}{\normalsize\sc}}
\makeatother

\newcommand{\nc}{\newcommand}
\newcommand{\rnc}{\renewcommand}

\nc{\chap}[1]{{\clearpage}%
\begin{center}%
{\noindent\underline{\large\sc #1}}{\addcontentsline{toc}{section}{#1}}%
\end{center}%
{\vspace*{0.3cm}}}

\nc{\subs}[1]{{\vspace*{0.2cm}}%
{\noindent\underline{\small\sc
#1}}%
{\vspace*{0.2cm}}}

\nc{\be}{\begin{equation}}
\nc{\ee}{\end{equation}}
\nc{\bea}{\begin{eqnarray}}
\nc{\eea}{\end{eqnarray}}

\nc{\trac}[2]{{\textstyle\frac{#1}{#2}}}

\nc{\ex}[1]{\mbox{e}^{\,\textstyle#1}}

\nc{\CC}{\Bbb{C}}
\nc{\HH}{\Bbb{H}}
\nc{\PP}{\Bbb{P}}
\nc{\RR}{\Bbb{R}}
\nc{\ZZ}{\Bbb{Z}}
\nc{\II}{\Bbb{I}}
\nc{\EE}{\Bbb{E}}
\nc{\TT}{\Bbb{T}}
\nc{\DD}{\mathrm{I}\!\mathrm{D}}

\rnc{\d}{\delta}
\nc{\eps}{\epsilon}
\nc{\om}{\omega}
\nc{\symx}{\circledS}
\newsymbol\smallsmile 1360
\newsymbol\smallfrown 1361
\nc{\ad}{\mathop{\mbox{ad}}\nolimits}
\nc{\tr}{\mathop{\mbox{tr}}\nolimits}
\nc{\Tr}{\mathop{\mbox{Tr}}\nolimits}
\nc{\Det}{\mathop{\mbox{Det}}\nolimits}
\rnc{\det}{\mathop{\mbox{det}}\nolimits}
\nc{\rk}{\mathop{\mbox{rk}}\nolimits}
\nc{\del}{\partial}
\nc{\diag}{\mathop{\mbox{diag}}\nolimits}
\nc{\ra}{\rightarrow}
\nc{\Ra}{\Rightarrow}
\nc{\LRa}{\Leftrightarrow}
\nc{\lra}{\leftrightarrow}
\nc{\ot}{\otimes}
\rnc{\ss}{\subset}
\nc{\nul}{\noindent\underline}
\nc{\non}{\nonumber\\}
\nc{\mat}[4]{\left(\begin{array}{cc}#1&#2\\#3&#4\end{array}\right)}
\rnc{\lg}{\frak{g}}
\nc{\G}[3]{\Gamma^{#1}_{\;{#2}{#3}}}
\nc{\nam}{\nabla_{\mu}}
\nc{\nan}{\nabla_{\nu}}
\nc{\dx}{\dot{x}}
\nc{\tx}{\tilde{x}}
\nc{\dtx}{\dot{\tilde{x}}}
\nc{\te}{\tilde{e}}
\nc{\dte}{\dot{\tilde{e}}}
\nc{\dxl}{\dot{x}^{\la}}
\nc{\dxm}{\dot{x}^{\mu}}
\nc{\dxn}{\dot{x}^{\nu}}
\nc{\ddx}{\ddot{x}}
\nc{\ddxm}{\ddot{x}^{\mu}}
\nc{\ddxn}{\ddot{x}^{\nu}}
\nc{\dxi}{\dot{\xi}}
\nc{\ddxi}{\ddot{\xi}}
\nc{\lsf}{\ell_s^{\mathrm{eff}}}
\nc{\lpf}{\ell_p^{\mathrm{eff}}}
\nc{\sqg}{\sqrt{g^{11}}}

\nc{\bpm}{\begin{pmatrix}}
\nc{\epm}{\end{pmatrix}}

\nc{\red}[1]{{\color{red}#1}}
\nc{\dd}{d}

\nc{\N}{\mathcal{N}}
\nc{\V}{\mathcal{V}}

\begin{document}
\vspace*{1cm}
%\rightline{Version of \today}
\thispagestyle{empty}

\begin{center}
{\Large{\textsc{Horizon Shells:\\[.3cm]Classical Structure at the Horizon of a Black Hole}}}
%Is the Black Hole Horizon really featureless?}}}
\end{center}
\vspace{.5cm}
\begin{center}
\large{\textsc{Matthias Blau}${}^1$ and \large{\textsc{Martin O'Loughlin}}$^{2}$\\[.8cm]
$^{1}$Albert Einstein Center for Fundamental Physics\\
Institute for Theoretical Physics, University of Bern\\
Sidlerstrasse 5, 3012 Bern, Switzerland}\\[.5cm]
%\texttt{blau@itp.unibe.ch} (corresponding author)\\[.5cm]
$^{2}$ University of Nova Gorica\\
Vipavska 13,  5000 Nova Gorica, Slovenia
%\texttt{martin.oloughlin@ung.si}
\end{center}
\vspace{1cm}

We address the question of the uniqueness of the Schwarzschild black hole
by considering the following question: How many meaningful solutions of
the Einstein equations exist that agree with the Schwarzschild solution
(with a fixed mass $m$) everywhere except maybe on a codimension one
hypersurface? The perhaps surprising answer is that the solution is
unique (and uniquely the Schwarzschild solution everywhere in spacetime)
\textit{unless} the hypersurface is the event horizon of the Schwarzschild
black hole, in which case there are actually an infinite number of
distinct solutions. We explain this result and comment on some of the
possible implications for black hole physics.

\vfill
\begin{center}
\textit{Essay written for the Gravity Research Foundation 2016 Awards for Essays on Gravitation}
\end{center}
\vfill
\newpage

%\section{Introduction}

\setcounter{page}{1}
The Schwarzschild solution is arguably the single most important exact
solution of the Einstein field equations. In particular, it is the
prototypical black hole metric, exhibiting an event horizon which
is a Killing horizon, and by the celebrated Israel theorem (1967)
\cite{Israel}, the first black hole uniqueness (``no hair'') theorem,
under certain regularity conditions it is the unique static black hole
solution of the vacuum Einstein equations.

While these classical aspects of the Schwarzschild solution have thus
been completely understood for a long time, the deep puzzles raised 
by the discovery of Hawking radiation and the thermodynamical properties
of black holes continue to fuel heated debates regarding the nature
of black hole microstates and the physics at or in the vicinity of a 
black hole horizon.

Many of these considerations suggest that at the quantum level
the black hole horizon may be a special place not just globally
%where it has 
%a coordinate invariant significance 
%as a boundary between observers who can in principle remain outside 
%and others who can never leave the region interior region, 
but even
locally in the space-time. In light of this let us take a step back and ask the seemingly
innocuous purely classical

\begin{description}
\item[Question:] How many meaningful solutions of the Einstein equations exist that
agree with the Schwarzschild solution (with a fixed mass $m$) 
everywhere except perhaps on a codimension one
hypersurface?
\end{description}
 
This question is a special case of a general problem analysed in \cite{MMshells}, 
and, as we will explain, has the following (perhaps somewhat surprising)

\begin{description}
\item[Answer:] The solution is unique (and uniquely the Schwarzschild solution
everyhwere in spacetime) \textit{unless} the hypersurface is the event horizon
of the Schwarzschild black hole, in which case there are actually an infinite
number of distinct solutions.
\end{description}

These solutions (almost all of which are new) are described in detail
in \cite{MMshells}. In particular, this result shows that there can be
a very rich structure at the horizon of a black hole already at the
classical level, with potential implications for various aspects of 
black hole physics.

%The purpose of this essay is to explain this result in simple terms and
%to explore some of its possible consequences and implications.

%\section{Soldering Freedom and Schwarzschild Horizon Shells}

To set the stage, we need to explain what we mean by ``meaningful
solutions of the Einstein equations except on a codimension one 
hypersurface'' in the above question. Thus let us imagine that we have 
two solutions of the Einstein equations
that we want to glue (or solder) along a common boundary hypersurface 
$\N$ to create a joint solution of the Einstein equations. Then the primordial 
(and obviously necessary) mathematical consistency condition, expressing the
fact that at each point in the extended space-time there should be a
unique metric, is the \textit{Israel junction condition} that the two metrics
should induce the same metric on $\N$. 
%It is common to write this condition as
%\be
%[g_{ab}]\equiv g_{ab}^+ - g_{ab}^- = 0\;\;,
%\ee
%expressing the statement that there is no jump in the induced metric
%across the shell.

It turns out that this necessary mathematical condition is also sufficient
to arrive at a physically meaningful solution of the Einstein equations,
namely with non-singular particle propagation across the hypersurface, and with
a delta-function localised contribution to the energy-momentum
tensor,
\be
G_{\alpha\beta} = \text{(bulk terms)} + S_{\alpha\beta}\delta_{\N}
\ee
(and/or to the Weyl tensor). The generic solution of this type
can be interpreted as describing a shell of matter on $\N$ together with
impulsive gravitational radiation in the case that $\N$ is null. 
%(and/or
%impulsive gravitational radiation) on $\N$ (this latter possibility
%evidently only arises in the case that $\N$ is null). 
The general formalism of thin shells
in general relativity (see e.g.\ \cite{poisson1} for a nice account,
and \cite{BI, BH} for further details regarding the more subtle case of
null shells) provides an algorithm to determine the shell energy
momentum tensor $S_{\alpha\beta}$ and other intrinsic physical properties
of the shell from the jumps in the normal derivatives of the metric across
the shell.

So far this is completely general, and for generic hypersurfaces
(including all timelike and spacelike hypersurfaces but also generic null
hypersurfaces) such a soldering, when it exists, is unique, i.e.\ there
is (up to isometries) 
%at most a (then essentially, i.e.\ up to isometries) 
a unique solution to the Israel junction condition. 
In particular, to return to the question raised at the
beginning of this essay, in all these cases there is a unique solution
to the Einstein equations that can be obtained by soldering two equal
mass Schwarzschild black hole solutions across such a hypersurface, namely
the Schwarzschild solution without a shell (equivalently, philosophical
questions aside, with an empty shell).

%In particular, to return to the question raised at the
%beginning of this essay, in all these cases there is a unique solution
%to the Einstein equations that can be obtained by soldering two equal
%mass Schwarzschild black hole solutions across such a hypersurface, namely
%the Schwarzschild solution without a shell (equivalently, philosophical
%questions aside, with an empty shell).

However, it was already mentioned in \cite{BI} that for the particular case of Killing
horizons of static black holes there is considerable freedom in the way that the
two geometries are attached,  and isolated examples of this phenomenon
are already present in the literature (e.g.\ the Dray 't Hooft
shell on the horizon of a Schwarzschild black hole \cite{DT2}). Intuitively
(but not completely correctly) this freedom can be atttributed to the 
fact that any null matter placed on the horizon will be infinitely redshifted
relative to any point a finite distance from the horizon and will, 
in particular, have no impact on quantities like the ADM mass of the
solution (unlike matter on a collapsing non-horizon shell, say).

Analysing this issue more systematically, in \cite{MMshells} we found
that there is a (continuously) infinite number of distinct ways to
solder two geometries together along a null hypersurface whenever
the induced metric on the null hypersurface
is invariant under translations along its null generators.
This condition is of course satisfied by Killing horizons of
stationary black holes, but also by Rindler horizons, and more generally
by other quasi-local notions of horizons such as non-expanding horizons
and isolated horizons (see e.g.\ \cite{ih1} for a review). 

A concrete illustration of this is provided by the Schwarzschild metric in
(ingoing) Eddington-Finkelstein coordinates, 
\be
ds^2 = -f(r) dv^2 + 2 dvdr + r^2 d\Omega^2\quad,\quad f(r) = 1-\frac{2m}{r}\;\;.
\ee
The metric induced on a hypersurface of the form $r=r(v)$ (say), will 
depend non-trivially on $v$ and/or $dv$, unless one makes the special horizon choice 
$r(v)=2m$. In this case, and this case only, the induced (degenerate) metric 
\be
ds^2|_{r=2m} = (2m)^2 d\Omega^2
\ee
is manifestly invariant under arbitrary transformations 
\be
v \ra F(v,y^A)
\ee
of the null coordinate $v$ ($y^A$ are the angular coordinates),
extended to or induced by a suitable coordinate transformation on one side of the 
shell. This provides one with an infinite number of distinct solutions to the Israel junction condition,
and
any such soldering transformation can be regarded as generating a shell from the empty shell (pure
Schwarzschild solution). This shell will be non-trivial unless the soldering transformation 
is a pure isometry, 
%transformation, 
i.e.\ a constant shift of $v$. 

The resulting shells will in general carry a conserved (null) matter energy momentum
tensor (composed of energy density $\mu$, energy currents or momentum
densities $j^A$ and pressure $p$), as well as impulsive gravitational
waves travelling along the shell, and following the null shell algorithm
these quantities can be expressed explicitly in terms of the 
first and second derivatives of $F$. 
For example the null energy density of the shell
$\mu$ is given by 
\be
\mu = -\frac1{8m\pi F_v}\left(\ex{-F/4m}\Delta^{(2)}\ex{F/4m} + F_v - 1\right)
\ee
(with $\Delta^{(2)}$ the Laplacian on the 2-sphere).
%and the quantities $\mu,j^A,p$ are linked by a conservation law of the form
%\be
%\partial_v \mu + \nabla^{(2)}_A j^A + \ldots = 0 
%\ee
%(here $\nabla^{(2)}$ and $\Delta^{(2)}$ denote the covariant
%derivative and Laplacian with respect to the metric on the unit 2-sphere).
 
To summarise, for any function (soldering transformation) $F(v,y^A)$ we obtain a 
solution to the Einstein equation describing a shell living on the horizon of a
Schwarzschild 
black hole. They generalise the known solutions of this kind in the following way:
\begin{itemize}
\item
These solutuions can be considered as significant generalisations of the
Dray 't Hooft null shell \cite{DT2}, to which they reduce for a special soldering
transformation which is a constant shift of the Kruskal-Szekeres coordinate
$V$, 
\be
V= \ex{v/4m} \ra V + b \quad\LRa\quad F(v,y^A) = 4m \log \left(\ex{v/4m}+b\right)
\ee
(this leads to a shell with constant energy density $\mu$ and zero pressure, currents and 
gravitational wave components). 
\item
As all of the solutions are non-singular
on the shell (i.e.\ do not have any point particle singularities),
they also provide one with a wide array of smoothed out versions and generalisations of
Dray 't Hooft impulsive gravitational waves \cite{DT1}. 
\item Finally these solutions reduce to those discussed by Barrabes and Israel in \cite{BI} in the special
case $F=F(v)$.
\end{itemize}

Whatever the ultimate significance or role of these solutions may be, we
believe that it is at the very least good to be aware of their existence,
and we conclude this essay with some remarks on the possible implications
of this result for black hole physics.

%\section{Speculations? Outlook? Whatever ...}

\begin{enumerate}

\item The most conservative interpretation is perhaps that these shells
are bookkeeping devices that (in the spirit of the method of images or 
perhaps the membrane paradigm)
encode faithfully the effects of coordinate transformations performed
on one side of the horizon. Performing such one-sided coordinate transformations 
may be relevant to a complete description of the physics of black holes for observers 
in the exterior region as required by the principle of black hole complementarity. 

\item
A class of soldering transformations that may be of particular interest, 
especially in light of the observation in \cite{andy2} that black holes
must carry supertranslation hair, and the subsequent
Hawking - Perry - Strominger proposal
\cite{Hawking1, Hawking2} (cf.\ also the analysis by Comp\`ere and Long \cite{CL}) 
are the horizon analogues of BMS supertranslations at $\mathcal{I}^+$, (infinitesimally)
of the form 
\be
v\ra v + T(x^A)\;\;,
\ee
which are singled out by commuting with time ($v$) translations. For these one finds
that the corresponding shells have zero pressure, but non-trivial $\mu$ and $j^A$ as
well as non-trivial impulsive gravitational shock waves travelling along the horizon.

\item As circumstantial evidence for the suggestion that these shells may be more
than an abstract bookkeeping device, we note that the Goldstone (-Bogoliubov) mode
of spontaneoulsy broken supertranslation invariance identified in \cite{Hawking2}
(cf.\ also \cite{ADGL}) is precisely the null energy density $\mu$ of the corresponding
shell, suggesting a concrete physical origin and interpretation of the null matter and radiation
propagating on the horizon shell.
\end{enumerate}

Clearly there are also many other issues, both classical and quantum, that remain to be explored.

\subsection*{Acknowledgements}

The research of MO was funded by the Slovenian Research Agency.
MO acknowledges financial support from the
EU-COST Action MP1210 ``The String Theory Universe'' and from the
Albert Einstein Center for Fundamental Physics, Bern. 
The work of MB is partially supported through the NCCR SwissMAP (The Mathematics
of Physics) of the Swiss Science Foundation.

\rnc{\Large}{\normalsize}

\end{document}